\documentclass[fleqn,usenatbib,useAMS]{mnras}

\usepackage{newtxtext,newtxmath}

\usepackage[T1]{fontenc}

\DeclareRobustCommand{\VAN}[3]{#2}
\let\VANthebibliography\thebibliography
\def\thebibliography{\DeclareRobustCommand{\VAN}[3]{##3}\VANthebibliography}

\usepackage{graphicx}	
\usepackage{amsmath}	
\usepackage{orcidlink}
\usepackage{url}
\usepackage{pdflscape}
\usepackage{threeparttable}
\usepackage{caption}
\usepackage{pgffor}
\usepackage{lineno}

\newcommand{\ie}{i.e., }
\newcommand{\eg}{e.g., }

\newcommand{\teff}{T_{\rm eff}}
\newcommand{\kms}{km\,s$^{-1}$}

\title[NIR spectra of CP stars and kilonovae]{Near-infrared spectra of chemically peculiar stars and applications to kilonova spectra}

\author[N. Domoto et al.]{
Nanae Domoto\,\orcidlink{0000-0002-7415-7954},$^{1}$\thanks{E-mail: ndomoto@g.ecc.u-tokyo.ac.jp}
Masaomi Tanaka\,\orcidlink{0000-0001-8253-6850},$^{2,3}$
Wako Aoki\,\orcidlink{0000-0002-8975-6829},$^{4,5}$
Salma Rahmouni\,\orcidlink{0009-0002-1232-243X},$^{2}$
\newauthor
Tomoyuki Kudo\,\orcidlink{0000-0002-9294-1793},$^{6}$
Hiroki Harakawa\,\orcidlink{0000-0002-7972-0216},$^{7}$
Teruyuki Hirano\,\orcidlink{0000-0003-3618-7535},$^{8,4}$
Takayuki Kotani\,\orcidlink{0000-0001-6181-3142},$^{8,4,5}$
\newauthor
Takashi Kurokawa,$^{9}$
Masayuki Kuzuhara\,\orcidlink{0000-0002-4677-9182},$^{8,4}$
Jun Nishikawa\,\orcidlink{0000-0001-9326-8134},$^{8,4}$
Takuma Serizawa,$^{9,4}$
\newauthor
Motohide Tamura\,\orcidlink{0000-0002-6510-0681},$^{10,8}$
and Akitoshi Ueda$^{8,4,5}$
\\
$^{1}$Research Center for the Early Universe, Graduate School of Science, The University of Tokyo, 7-3-1 Hongo, Bunkyo-ku, Tokyo 113-0033, Japan\\
$^{2}$Astronomical Institute, Tohoku University, Aoba, Sendai 980-8578, Japan\\
$^{3}$Division for the Establishment of Frontier Sciences, Organization for Advanced Studies, Tohoku University, Sendai 980-8577, Japan\\
$^{4}$National Astronomical Observatory of Japan, 2-21-1 Osawa, Mitaka, Tokyo 181-8588, Japan\\
$^{5}$Astronomical Science Program, The Graduate University for Advanced Studies, SOKENDAI, 2-21-1 Osawa, Mitaka, Tokyo 181-8588, Japan\\
$^{6}$Subaru Telescope, National Astronomical Observatory of Japan, 650 North A'ohoku Place, Hilo, HI 96720, USA\\
$^{7}$Department of Science, National Museum of Nature and Science, 4-1-1 Amakubo, Tsukuba, Ibaraki 305-0005, Japan\\
$^{8}$Astrobiology Center, 2-21-1 Osawa, Mitaka, Tokyo 181-8588, Japan\\
$^{9}$Institute of Engineering, Tokyo University of Agriculture and Technology, 2-24-16 Nakacho, Koganei, Tokyo 184-8588, Japan\\
$^{10}$Department of Astronomy, Graduate School of Science, The University of Tokyo, 7-3-1 Hongo, Bunkyo-ku, Tokyo 113-0033, Japan
}

\date{Accepted XXX. Received YYY; in original form ZZZ}

\pubyear{2026}

\begin{document}
\label{firstpage}
\pagerange{\pageref{firstpage}--\pageref{lastpage}}
\maketitle

\begin{abstract}
Kilonova spectra provide information on rapid neutron-capture nucleosynthesis in neutron star mergers. However, it remains challenging to identify individual elements in kilonova spectra, primarily due to incomplete atomic data for heavy elements at near-infrared wavelengths. 
Recent work demonstrated that spectra of chemically peculiar stars offer excellent laboratory for studying kilonova spectra. 
In this paper, to expand our understanding of strong transitions of heavy elements at near-infrared wavelengths, we present high-resolution near-infrared spectra of 14 chemically peculiar stars covering a range of effective temperatures, surface gravities, and abundance patterns, obtained with Subaru/IRD. 
The \ion{Sr}{ii}, \ion{Ce}{iii}, and \ion{Gd}{iii} lines, previously suggested to be among the strongest transitions, are detected in the spectra of most of our targets, confirming their intrinsic importance. In addition, \ion{Eu}{ii} and \ion{Nd}{iii} lines are detected in several stars.
By comparing observed equivalent widths and abundance ratios, 
we show that under a highly heavy-element-enriched abundance pattern, \ion{Eu}{ii} and \ion{Nd}{iii} lines may contribute to kilonova spectral features comparably to \ion{Sr}{ii}.
Importantly, we find no unidentified lines systematically stronger than these transitions despite the larger stellar sample. These results further support the primary importance of the \ion{Sr}{ii} and \ion{Ce}{iii} transitions and provide new insights into the potential contribution of \ion{Eu}{ii} and \ion{Nd}{iii}.
\end{abstract}

\begin{keywords}
line: identification -- stars: chemically peculiar -- neutron star mergers -- atomic data
\end{keywords}



\section{Introduction}
\label{sec:intro}
Coalescence of binary neutron stars has been considered a promising site of $r$-process nucleosynthesis \citep[e.g.,][]{LS1974, Eichler1989, Freiburghaus1999, Goriely2011, Wanajo2014}.
Associated with the detection of gravitational waves (GWs) from a neutron star merger GW170817 \citep{Abbott2017a}, its electromagnetic counterpart AT2017gfo was observed at the ultraviolet, optical, and infrared wavelengths \citep{Abbott2017b}. The observational properties of AT2017gfo are consistent with those expected for so-called kilonova \citep{Li1998, Metzger2010}, providing strong evidence that neutron star mergers are the site of $r$-process nucleosynthesis \citep[see e.g.,][for reviews]{Metzger2020, Nakar2020, Margutti2021}.

For AT2017gfo, a series of optical to near-infrared (NIR) spectra have been obtained \citep[e.g.,][]{Pian2017, Smartt2017}.
To extract detailed information on $r$-process nucleosynthesis in neutron star merger ejecta, it is necessary to identify individual elements in kilonova spectra, as done for stellar spectra. Recent advances in the study of kilonova spectra have reported direct identification of elements (see Figure~\ref{fig:kn} below): \ion{Sr}{ii} or \ion{He}{i} in the 0.9\,$\umu$m absorption features \citep[atomic number $Z=38, 2$;][]{Watson2019, Domoto2021, Gillanders2022, Perego2022, Arya2026, Chiba2026}, \ion{Y}{ii} in the 0.6\,$\umu$m features \citep[$Z=39$;][]{Sneppen2023}, \ion{La}{iii} and \ion{Gd}{iii} in the 1.3\,$\umu$m features \citep[$Z=57, 64$;][]{Domoto2022, Rahmouni2025}, and \ion{Ce}{iii} in the 1.45\,$\umu$m features \citep[$Z=58$;][]{Domoto2022, Domoto2023,Tanaka2023}, as well as \ion{Te}{iii} in the 2.1\,$\umu$m emission features \citep[$Z=52$;][]{Hotokezaka2023}. 
Although no conclusive element identification has been made, the importance of Th ($Z=90$) in absorption lines, and Se ($Z=34$) and W ($Z=74$) in emission lines, has also been discussed \citep{Hotokezaka2022, Domoto2025}.

However, element identification in kilonova spectra remains challenging, primarily due to the lack of accurate atomic data for heavy elements at NIR wavelengths.
\citet{Domoto2022} constructed a ``hybrid'' line list by combining a theoretical line list from \citet{Tanaka2020} with an experimentally accurate line list. Although their data are calibrated for several species with strong important transitions, most of the data remain uncalibrated.\footnote{In this context, calibration means shifting the calculated energy levels and transition wavelengths to the experimentally accurate values.}
Recently, \citet{Floers2026} presented a calibrated line list for all singly and doubly ionized lanthanides based on theoretical calculations, but their transition probabilities still carry uncertainties that require experimental validation. 
Alternatively, \citet{Gillanders2024} and \citet{Rahmouni2025} used experimentally accurate data for low-lying energy levels and E1 selection rules to identify important transitions, with \citet{Gillanders2024} compiling a shortlist of transitions that may explain the observed features in the AT2017gfo spectra. 
Nevertheless, this approach relies on the completeness of experimental data, which may not be certain, and cannot determine strength of features without transition probabilities. 
Consequently, there are still no complete and accurate data at NIR wavelengths covering all heavy elements synthesized in neutron star mergers. It raises a question about the uniqueness of element identification and leaves the possibility that unknown transitions contribute to the NIR absorption features in kilonova spectra.

To address this situation, \citet{Tanaka2023} demonstrated that the atmospheres of chemically peculiar (CP) stars provide an excellent empirical laboratory for studying kilonova spectra.
CP stars are stars on or near main-sequence that exhibit abnormal abundance patterns. They are classified into several types, including non-magnetic AmFm that are underabundant in Ca and Sc, and magnetic ApBp stars that generally have strong magnetic fields of order kG and exhibit overabundances of heavy elements, especially for lanthanides \citep{Preston1974}. 
Among these, rapidly oscillating Ap (roAp) stars are known for exhibiting abundance discrepancies between singly and doubly ionized lanthanides, called the rare-earth element (REE) anomaly \citep{Ryabchikova2004, Ryabchikova2017}.
\citet{Tanaka2023} particularly discussed lanthanides using the Bp star HR~465, which has enhanced abundances of trans-iron elements and ionization states similar to those in neutron star merger ejecta. They showed that there are no other transitions with comparable strength to the \ion{Ce}{iii} lines at nearby wavelengths, one of the species reported in the spectra of AT2017gfo \citep{Domoto2022, Domoto2023}, which strengthens the identification of Ce in the kilonova spectra. 
However, because they studied only a single object, it remained unclear whether the dominance of \ion{Ce}{iii} is always the case.
In addition, their conclusion was restricted by a single set of stellar parameters, not perfectly identical to kilonova conditions.
Thus, a broader investigation of CP stars with various physical properties is warranted to comprehensively examine important NIR transitions of heavy elements.

In this paper, we investigate strong absorption lines by obtaining high-resolution NIR spectra of CP stars with a range of abundance ratios and atmospheric parameters. This allows us to expand our knowledge of strong transitions at NIR wavelengths for a better understanding of kilonova spectra. 
In Section~\ref{sec:obs}, we describe the sample selection and observations. We provide our analysis in Section~\ref{sec:ana}, and present the results in Section~\ref{sec:res}. 
The implications for kilonova spectra are discussed in Section~\ref{sec:discussion}.
Finally, we give our summary in Section~\ref{sec:conclusion}. 
In Appendix~\ref{sec:sr}, we report new transition wavelengths of \ion{Sr}{ii} measured through our analysis.
Theoretical behavior of the absorption lines used in the analysis is supplementarily discussed in Appendix~\ref{sec:cog}. The entire view of the observed spectra is presented in Appendix~\ref{sec:entire}.
Throughout this paper, wavelengths are given as those in vacuum.

\begin{table*}
\centering
\begin{threeparttable}
	\caption{Parameters of the sample CP stars. Primary references for each star are provided: most atmospheric parameters are based on \citet{Ghazaryan2018} and references therein, while some atmospheric values as well as photometric/magnetic rotational variation period $P$ for ApBp stars are collected from other literature. Mark $\dag$ denotes the value derived in this work. Spectral types are taken from \citet{Renson2009} or SIMBAD.}
	\label{tab:params}
	\begin{tabular}{llccccccccc}\hline
		Star & Other ID & Spec. type & $\teff$ &      $\log g$      & [Fe/H] & $\xi^{\rm a}$ & $ v\sin i^{\rm b}$  & $P$  & Note & Refs. \\
		        &                   &                &     (K) & (cm\,s$^{-1}$) &             & (\kms) &   (\kms)   &  (day)  &        & \\  \hline
	    	HD~965       &                       & A8p SrEuCr                 & 7400  & 4.0 & $-0.50$   & 0.0   & 3.0                     & 6030      & roAp?$^*$ & 1--3 \\
		HD~5797     &                       & A0p SrEuCr                 & 8900  & 3.4 & 1.66        & 0.7   & 3.0                     & 68.05     & non-roAp  & 4, 5 \\
		HD~8441     &                       & A2p Sr                          & 9130 & 3.4 & 0.72        & 0.3   & 2.5                     & 69.51      & non-roAp, SB1  & 6, 7 \\
		HD~9996     & HR~465          & B9p CrEuSi                & 11000 & 4.0 & 0.92       & 2.0   & 2.0                     & 7850       &   SB1      & 8 \\
		HD~22316   & HR~1094        & B9p CrHgSi                & 12000 & 4.2 & 0.96       & 0.0   & 17.0                   & 2.977     &                & 9 \\
		HD~176232 & 10 Aql             &  A6p Sr                       & 7550   & 4.0 & 0.34       & 2.0   & 4.0                     & very long & roAp     & 10--12 \\
		HD~188041 & HR~7575        & A6p SrEuCr                & 8770  & 4.2  & 1.14       & 1.75 & 2.0                    & 223.83    & non-roAp & 13, 14 \\
		HD~203932 &                        & A5p SrEu                    & 7450  & 4.3 & 0.12       & 0.6   & 12.5                   & 6.44       & roAp       & 15--17 \\
		HD~225914 & KIC 4768731  & A5p SrCrEu                & 8100  & 4.0 & 0.29       & 0.5   & 14.8                   & 5.21       & roAp       & 18 \\
		HD~182564 & $\pi$ Dra        & A0III                           & 9125  & 3.8 & 0.33       & 3.5    & 25.0$^{\dag}$   & -             &                & 19 \\
		HD~187254 & KIC 8703413 & kA5hA5mF2V             & 8400  & 3.8 & 0.53       & 2.7   & 14.0                   & -             &   SB1      & 20 \\
		HD~189849 & 15 Vul            & A4III                            & 7850  & 3.7 & $-0.10$ & 4.0   & 13.0                    & -             &               & 21 \\
		HD~190165 & KIC 9117875 & kA3hF0.5mF3(III)Am  & 7300  & 3.8 & 0.45       & 1.9   & 61.0                   & -             &               & 20 \\
		HD~225463 & KIC 5633448 & kA3hA4mA7VAm       & 8300  & 3.8 & 0.07       & 2.6   & 13.0                   & -             &                & 20 \\ \hline
	\end{tabular}
	\begin{tablenotes}
		\item[a] Microturbulent velocity.
		\item[b] Rotational velocity.
		\item[*] While all properties are similar to those of roAp stars, the rapid oscillation has not been found \citep{Elkin2005}.
		\item[]{\it References:} (1) \citet{HD965_1}; (2) \citet{Mathys2019}; (3) \citet{HD965_2}; (4) \citet{Semenko2011}; (5) \citet{HD5797_rot}; (6) \citet{Titarenko2012}; 
		(7) \citet{HD8441_rot}; (8) \citet{Nielsen2020}; (9) \citet{Nielsen2000}; (10) \citet{Ryabchikova2000}; (11) \citet{Elkin2008}; (12) \citet{Nesvacil2013}; 
		(13) \citet{HD188041_rot}; (14) \citet{Romanovskaya2019}; (15) \citet{Gelbmann1998}; (16) \citet{Ryabchikova2000}; (17) \citet{HD203932_rot}; 
		(18) \citet{Smalley2015}; (19) \citet{Adelman1997}; (20) \citet{Niemczura2015}; (21) \citet{15Vul}
	\end{tablenotes}
\end{threeparttable}
\end{table*}

\begin{figure*}
	\includegraphics[width=\linewidth]{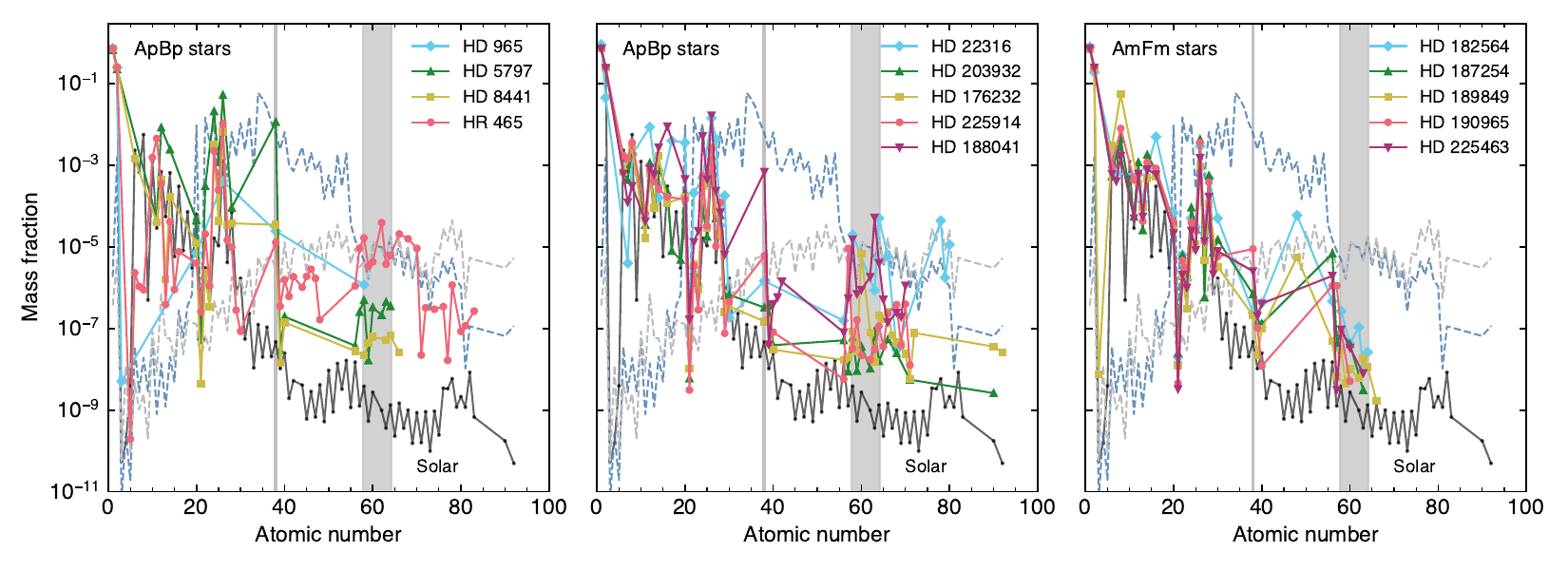}
	\caption{Elemental abundances of the sample ApBp stars (left and middle panels) and AmFm stars (right panel), compared with the solar values \citep[black;][]{Asplund2009}. The vertical axis shows the mass fraction, obtained by normalizing the available abundance measurements to unity, based on \citet{Ghazaryan2018} and references therein, together with the Sr abundances estimated in this work (Section~\ref{sec:ana}). The dashed curves represent example of nucleosynthesis model abundances from numerical simulations of neutron star mergers for both the dynamical (grey) and disk (blue) ejecta \citep[model SFHo-120-150;][]{Fujibayashi2023}, normalized to the Eu abundance ($Z=63$) of HR~465. Elements relevant for our discussions, Sr ($Z=38$) and a subset of lanthanides ($Z=58$--64), are highlighted by grey shading.
	}
	\label{fig:abun}
\end{figure*}

\section{Observations}
\label{sec:obs}
The sample stars were selected from the catalogue of CP stars compiled by \citet{Ghazaryan2018} based on the following criteria: (1) the object is listed as a single star\footnote{We note that a few of the selected stars are in fact known single-lined spectroscopic binaries (SB1; see Table~\ref{tab:params}); however, since flux contamination from their companions is negligible, the targets are effectively treated as single stars.}; (2) its effective temperature ($\teff$) and surface gravity ($\log g$) are known; and (3) its elemental abundances, specifically metallicity ${\rm [Fe/H]}$\footnote{${\rm [X/Y]}=\log_{10}(N_{\rm X}/N_{\rm Y})-\log_{10}(N_{\rm X}/N_{\rm Y})_{\odot}$, where $N_{\rm X}$ and $N_{\rm Y}$ are the abundances of elements X and Y, respectively.} and ${\rm [Ce/H]}$, have been measured from optical spectra. 
Considering visibility, our final sample consists of nine ApBp stars, including the previously studied star HR~465 \citep{Tanaka2023}, and five AmFm stars.
Their atmospheric parameters are summarised in Table~\ref{tab:params}.

Given the $\teff$ and $\log g$ values in Table~\ref{tab:params}, heavy elements in these atmospheres are predominantly singly or doubly ionized. 
Typical ionization states are expected to be similar to that of neutron star merger ejecta a few days after the merger, ranging from regimes where doubly ionized species dominate to where singly and doubly ionized states coexist \citep{Tanaka2023}. 
Notably, since CP-star photospheres are hotter than in kilonovae, more transitions can be observed in CP-star spectra for a given ion. This enables a comprehensive study of strong heavy-element transitions, although we note that the apparent ionization degrees in roAp stars can deviate from a homogeneous atmosphere assumption.

The abundances of elements for the sample CP stars are plotted in Figure~\ref{fig:abun} (see Section~\ref{sec:ana}), compared with the solar values \citep{Asplund2009}. Example of nucleosynthesis models based on numerical neutron star merger simulations is also shown as dashed curves \citep[model SFHo-120-150;][]{Fujibayashi2023} arbitrarily scaled for comparison. It can be seen that the significant enhancements of heavy elements in CP-star atmospheres produce abundance patterns that broadly resemble those of neutron star merger ejecta.

We note that these stars are used only as proxies of metal-rich atmospheres, not as indicators of intrinsic nucleosynthesis patterns. 
The significant enhancements of heavy elements observed in CP stars are caused by atomic diffusion \citep{Michaud1970}, and their bulk abundance patterns may differ from those observed on the surface. 
Also, because BAF-type CP stars are relatively young and metal-rich, their components experienced many nucleosynthesis events. 
Nevertheless, although the underlying physical mechanisms are different, these stars offer unique macroscopic insights into the complex spectral properties governing heavy element-rich merger ejecta.

NIR spectra of the sample stars were obtained with InfraRed Doppler (IRD) at the Subaru Telescope \citep{Tamura2012, Kotani2018} on UT 2024 October 9. 
The spectra cover the $Y$, $J$, and $H$ bands with a spectral resolution of $R\sim70000$. 
This wavelength range covers the transitions responsible for the 0.9\,$\umu$m, 1.15\,$\umu$m, and 1.45\,$\umu$m features observed in the AT2017gfo spectra (see Figure~\ref{fig:kn}). 
Three telluric standard stars (HR 8489, HD~178207, and HD~192538), located close to the targets on the sky, were also observed. 
Typical single exposure times range from 20 to 540\,s for the targets with $J$-band magnitudes of $\approx4.5$--8.7.

Data reduction was performed using \textsc{PyIRD} \citep{pyird} version 1.1.0, a Python-based open-source data reduction tool for IRD, which handles detector noise removal, flat-fielding, one-dimensional spectral extraction, and wavelength calibration using a Th-Ar lamp. 
Typical accuracy of wavelength calibration is $< 0.1$\,{\AA} and at most 0.3\,{\AA} (see also Section~\ref{sec:ana}).
To account for night-glow contamination, particularly in the $H$ band, emission lines were simply masked using the list of \citet{OHlines}.
For targets with multiple exposures, individual one-dimensional spectra were averaged into a single spectrum and then normalized by continuum fitting. 
The final signal-to-noise ratio ranges from $\approx90$ to 400 per pixel, depending on the star and the echelle blaze position. 
Both the sample CP stars and telluric standard stars were reduced in the same way. Telluric absorption in the normalized CP-star spectra was corrected by dividing it by the normalized standard-star spectra with locally optimized scaling factors. 
Finally, Doppler corrections were applied using strong \ion{Si}{i} lines as well as the \ion{Sr}{ii} triplet and \ion{Ce}{iii} lines.

\begin{figure*}
	\includegraphics[width=\linewidth]{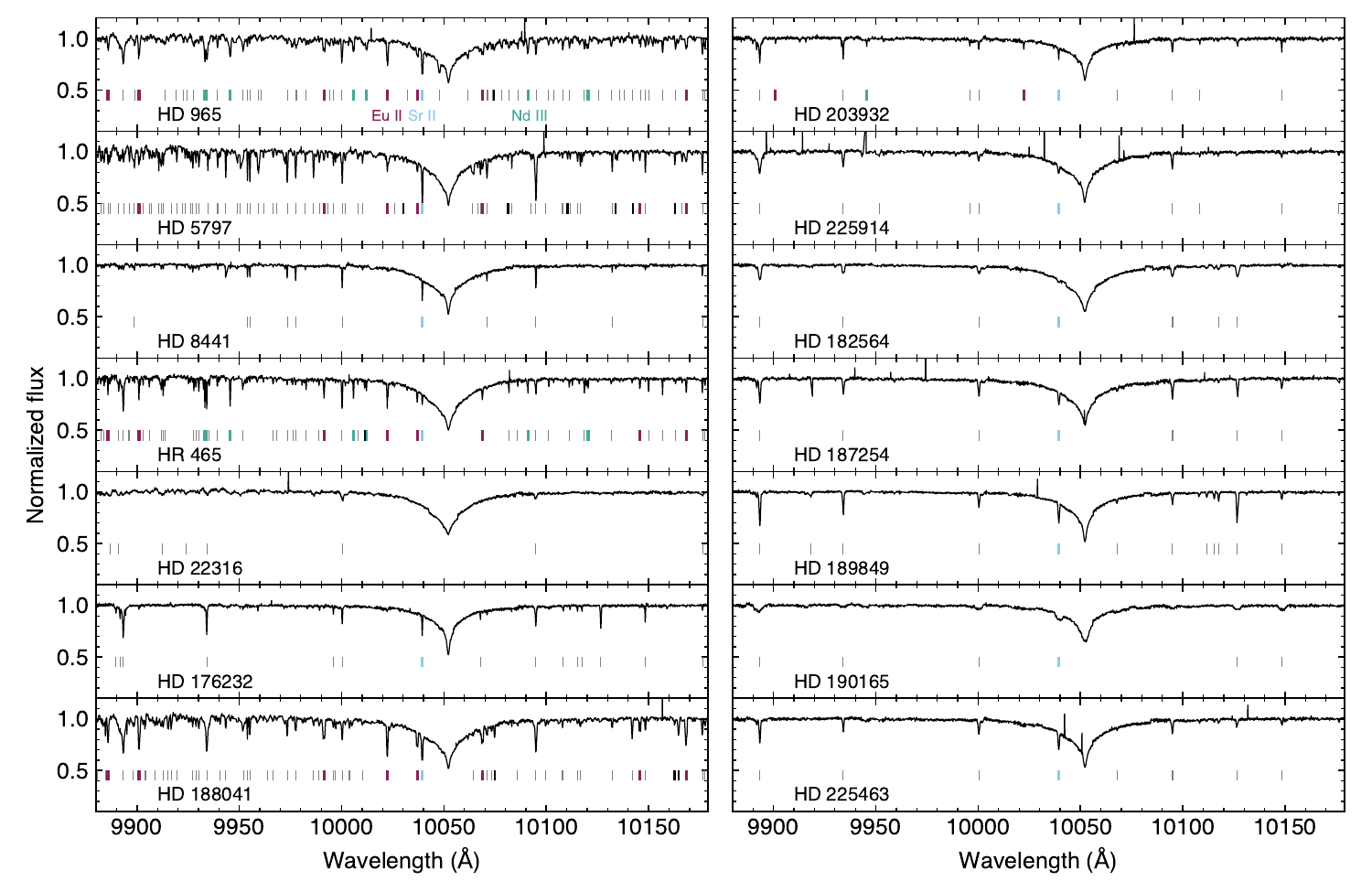}
	\caption{Example of spectra of the sample CP stars (object names are shown in each panel) obtained with Subaru/IRD in the $Y$ band, around the \ion{Sr}{ii} $\lambda$10039.41 line. Absorption lines with EWs $>16$\,m{\AA} are marked by vertical bars (blue: \ion{Sr}{ii}, wine: \ion{Eu}{ii}, rose: \ion{Ce}{iii}, green: \ion{Nd}{iii}, black: unidentified, and grey: others). See Appendix \ref{sec:entire} for the entire spectra.
	}
	\label{fig:yband}
\end{figure*}

\begin{figure*}
	\includegraphics[width=\linewidth]{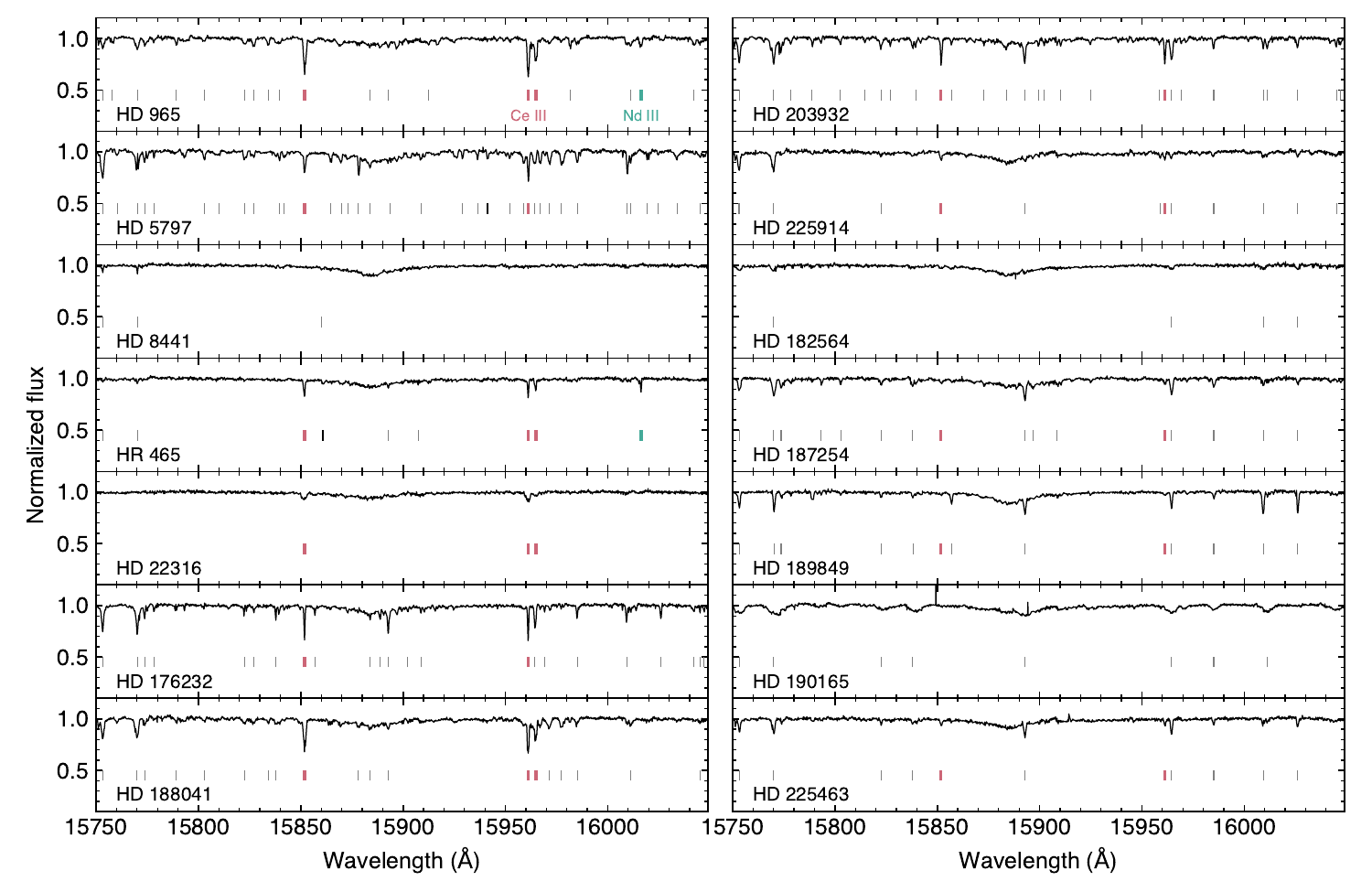}
	\caption{Same as Figure~\ref{fig:yband} but in the $H$ band, around the \ion{Ce}{iii} $\lambda$15851.88 and $\lambda$15961.16 lines. 
		      Absorption lines with EWs $>10$\,m{\AA} are marked by vertical bars.
		      }
	\label{fig:hband}
\end{figure*}

\section{Analysis}
\label{sec:ana}
In this section, we analyse the observed high-resolution NIR spectra of our sample CP stars. 
First, we detect strong absorption lines in the observed spectra. Wavelengths regions affected by strong telluric absorption, \ie those between the $Y$, $J$, and $H$ bands, are masked and excluded from line detection.

Second, we measure equivalent widths (EWs) of the detected absorption lines by fitting Gaussian or Voigt profiles.
For lines located near broad hydrogen features, the broad features are treated as a pseudocontinuum and locally flattened before measurement. 
In some ApBp stars, Zeeman splitting is detected for some strong lines due to strong magnetic fields. 
Since this is not essential for the purpose of this work, we smooth over the splitting and measure each line as a single absorption feature. 
Significant splitting observed in heavy-element lines, such as in HD~965 and HD~188041, may overestimate their EWs and affect the ranking (Section~\ref{sec:res}), but it does not affect our overall discussions.

Then, for the detected lines, we perform line identification.
Primary line identification is achieved by crossmatching the central wavelength of each detected absorption line against atomic line lists with a tolerance of 0.15\,{\AA}. 
We use the VALD line list \citep{Piskunov1995, Kupka1999, Ryabchikova2015} as a baseline, as well as the calibrated line list of \citet{Floers2026} for singly and doubly ionized lanthanides. 
For the \citet{Floers2026} list, we consider only transitions for which both energy levels are experimentally calibrated (flagged as 'xmatch'). 
If multiple candidate transitions are found for a detected line, we adopt the transition with the largest product of the weighted oscillator strength ($gf$-value) and the lower-level population ($e^{-E_{\rm lower}/kT}$). 
If multiple transitions of different elements have similar values, we adopt the one with the larger abundance. Lines for which no reasonable transition is found are labelled as unidentified.

To complement the wavelength crossmatching and resolve potential ambiguities, we compute synthetic model spectra for visual comparison. 
Spectral synthesis is performed with \textsc{Turbospectrum} \citep{Plez2012} through \textsc{iSpec} \citep{ispec1, ispec2}, assuming local thermodynamic equilibrium (LTE) and a one-dimensional plane-parallel atmosphere model ATLAS \citep{atlas}, with the stellar parameters primarily based on \citet{Ghazaryan2018} (Table~\ref{tab:params}).
For light elements, candidate lines are discarded if other lines of the same ion produce features inconsistent with the observed spectra. 
Conversely, because the $gf$-values in \citet{Floers2026} have not been experimentally assessed and therefore uncertain, we adopt a more flexible criterion for lanthanides: if the wavelength matches and the lower level of the transition is $E_{\rm lower}<5$\,eV where sufficient level population is expected at the $\teff$ range of our sample, the line is considered identified regardless of the model-data agreement.

For spectral synthesis, we need to assume elemental abundances. Beyond the REE anomalies in roAp stars, ApBp stars generally show abundance variations correlated with photometric and magnetic-field strength variations due to inhomogeneous abundance distribution on the stellar surfaces; the rotational period $P$ for our sample is listed in Table~\ref{tab:params}. Abundance variations in most of our sample are less than 1\,dex, but order-of-magnitude variations occur in HR~465, HD~22316, and likely in HD~965 \citep{Nielsen2000, Nielsen2020, Mathys2019}.\footnote{For HR~465, our observations were conducted at the phase when the lanthanide abundances reach a maximum ($\phi\approx0$). Compared with the spectrum in \citet{Tanaka2023}, the strength of \ion{Eu}{ii} and \ion{Nd}{iii} lines in our spectrum is significantly enhanced, likely due to this abundance variation, whereas that of \ion{Ce}{iii} and \ion{Sr}{ii} lines shows little change \citep[see also figure 5 of ][for the abundance variations of other ions]{Nielsen2020}. The phase for the HD~965 observation matches the secondary maximum of a Nd line EW \citep[$\phi\approx0.6$;][]{Mathys2019}, while HD~22316 was observed at $\phi\sim0.75$ which is near the phase at which lanthanide abundances likely reach a maximum \citep{Nielsen2000}.} 
However, we adopt the mean values obtained from the literature regardless of these variations (Figure~\ref{fig:abun}), assuming the solar values \citep{Asplund2009} when literature values are not available. For HD~965, where only Ce abundance is available among lanthanides, we scale the remaining lanthanides relative to the solar values using the logarithmic enhancement for Ce. These choices do not impact the complementary line identification and the global trend discussed in the paper. When we discuss specific ions in Section~\ref{sec:discussion}, the corresponding ionic abundances are adopted.

Although we do not require perfect agreement between models and observations, we find that the literature abundances for Sr poorly reproduce the observed \ion{Sr}{ii} triplet in many stars. This is likely because of some literature analyses relying on a single spectral line, as well as a significant saturation that complicates abundance measurements (see also Figure~\ref{fig:cog}), while literature abundances for other elements are mostly derived from multiple less-saturated lines.
Given the importance of \ion{Sr}{ii} to our discussions (Section~\ref{sec:discussion}), we re-estimate the Sr abundance using two of the triplet lines: $\lambda10039.41$ ($\log gf = -1.189$) and $\lambda10330.14$ ($\log gf = -0.247$) from the VALD line list. Modelled EWs are measured as done for the observed spectra, and the Sr abundance is iteratively adjusted to match the modelled EWs to the observed values. We adopt the mean Sr abundances that best reproduces the EWs of the observed lines (Table~\ref{tab:ews}), which should still be treated with caution, associated with a standard deviation of $\sim0.5$\,dex on average.

\begin{figure*}
	\includegraphics[width=\linewidth]{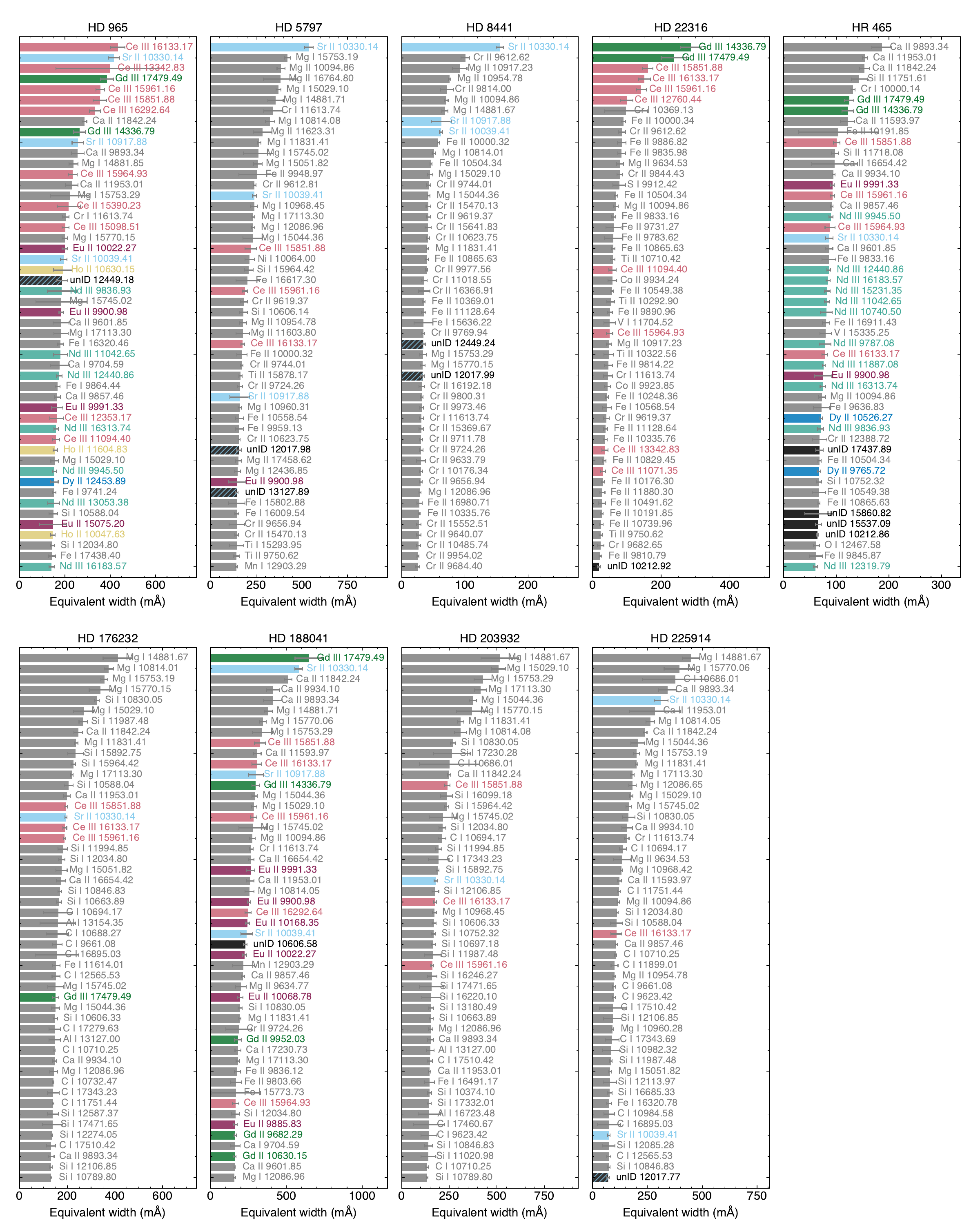}
	\caption{EWs of the 50 strongest lines in the NIR spectra of ApBp stars. Lines of elements heavier than Fe, which are abundant in neutron star merger ejecta, are shown in colours, while those of light elements are shown in grey. Unidentified lines that may appear in kilonova spectra are shown in black; those with hatches are likely \ion{Sr}{ii} (Appendix \ref{sec:sr}).
	}
	\label{fig:Ap}
\end{figure*}

\begin{figure*}
	\includegraphics[width=\linewidth]{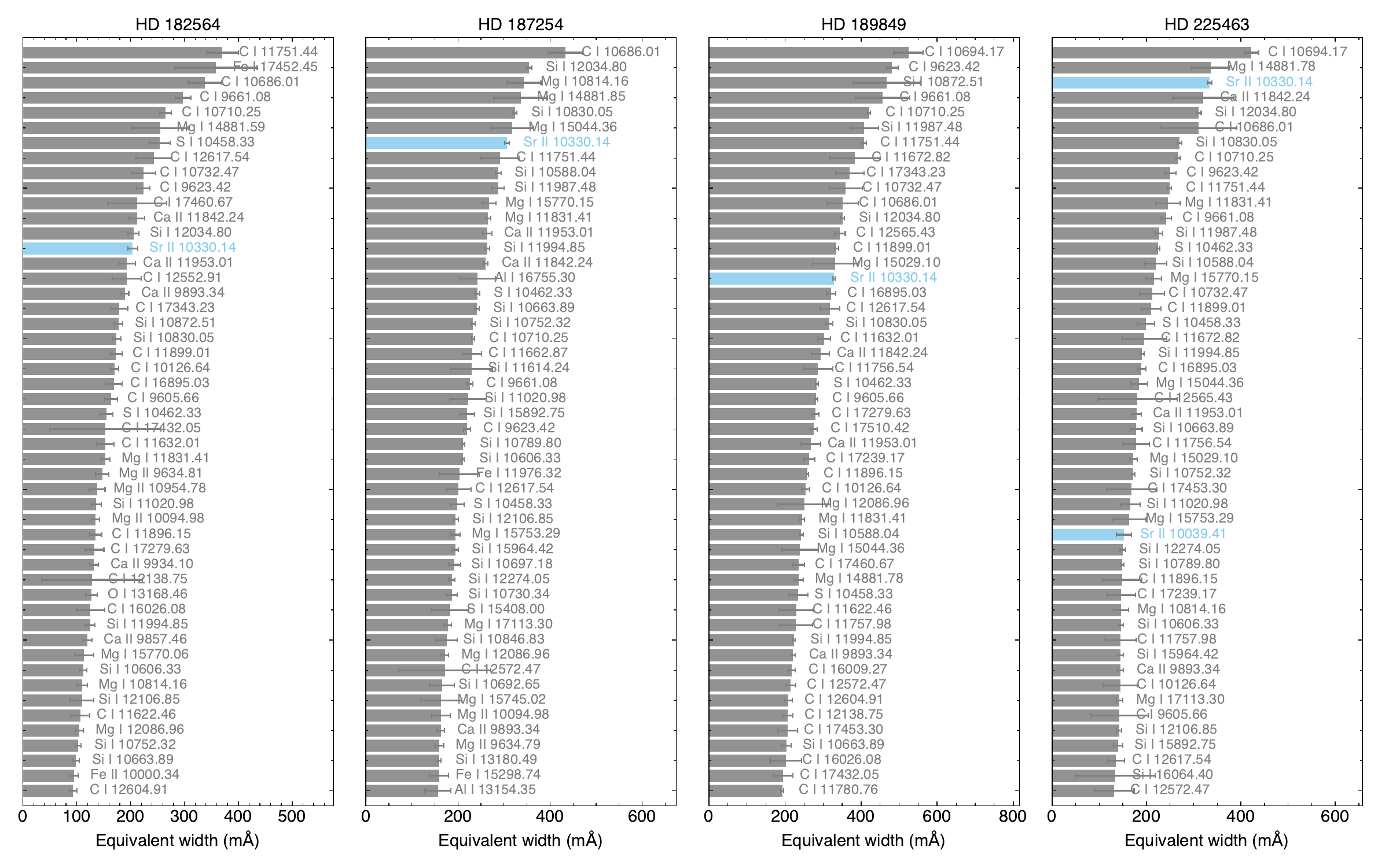}
	\caption{Same as Figure 4 but for AmFm stars except HD~190165.
	}
	\label{fig:Am}
\end{figure*}

\begin{figure*}
	\includegraphics[width=0.9\linewidth]{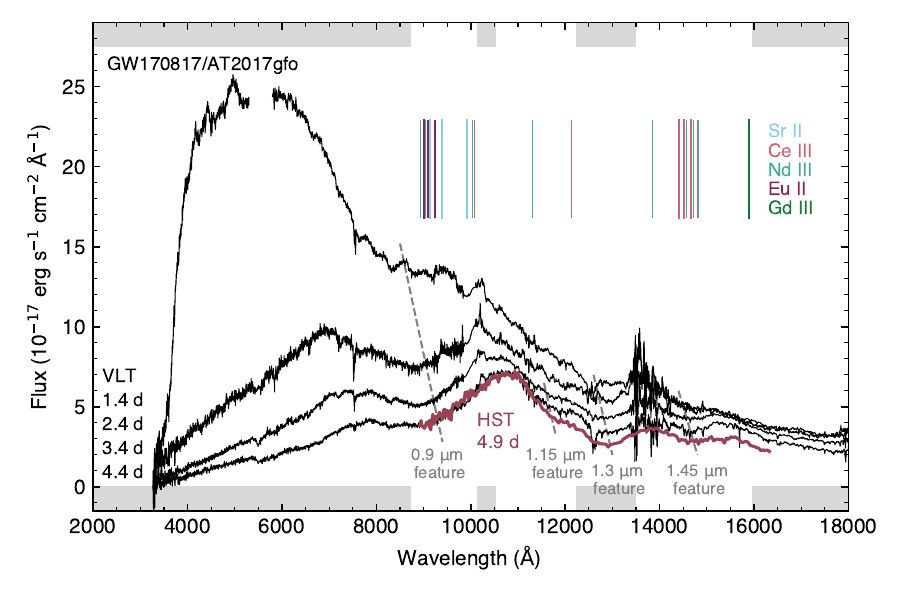}
	\caption{Spectra of AT2017gfo obtained with VLT \citep[1.4-4.4 days;][]{Pian2017, Smartt2017} and HST \citep[4.9 days;][]{Tanvir2017}. Thin and thick vertical bars mark the position of the strong heavy-element lines that are ranked in the top 50 for EWs in at least two or three individual stars, respectively, blueshifted by $v=0.1\,c$.
	Grey shaded region shows the wavelength range between the telluric windows for the stars (also blueshifted), where information on transitions are not available; the \ion{Gd}{iii} $\lambda$14336.79 line is omitted because it overlaps with a shaded region. Colours of the EWs are given according to the line identifications.
}
	\label{fig:kn}
\end{figure*}

\begin{figure*}
	\includegraphics[width=\linewidth]{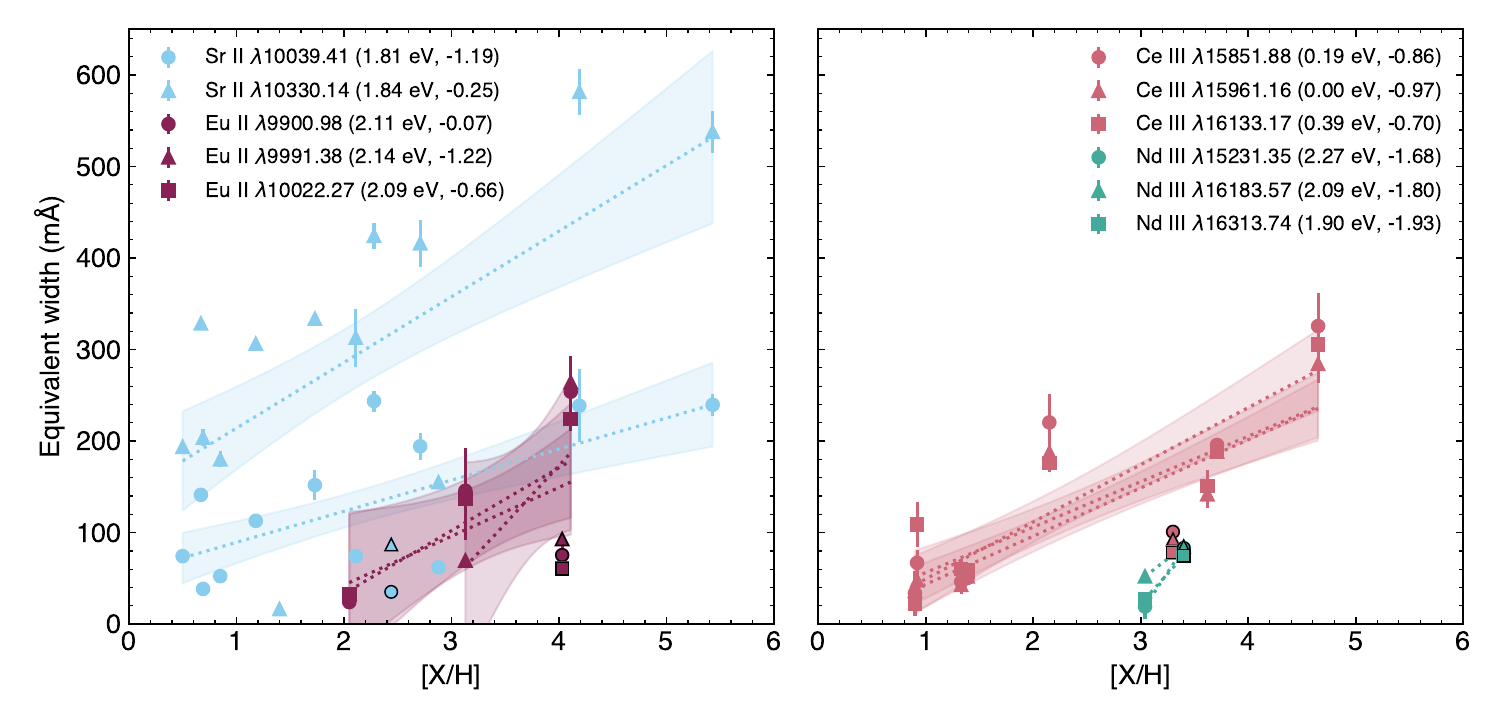}
	\caption{EWs of the selected lines indicated in the legend (wavelength, excitation energy, and $\log gf$) as a function of elemental abundance. Mean abundances measured using lines of the corresponding ionization states are adopted. EWs are not plotted if a measurement for that ionization state is not available in the literature. Error bars on the EWs show the fitting uncertainties. As a visual guide, the dotted lines show linear fits for each transition, with shaded regions indicating $1\sigma$ uncertainties. The dots for HR~465 are highlighted with borders (see the text). 
	}
	\label{fig:ew}
\end{figure*}

\section{Results}
\label{sec:res}
We detect absorption lines of heavy elements in all spectra. Examples are shown in Figure~\ref{fig:yband}--\ref{fig:hband}, and the entire spectra for all sample CP stars are presented in Figure~\ref{fig:entire}.

In most of our sample, we confirm the presence of key \ion{Sr}{ii} and \ion{Ce}{iii} lines. The \ion{Sr}{ii} triplet, $\lambda\lambda$10039.41, 10330.14, and 10917.88, is clearly detected in all targets except HD~22316, where only the strongest $\lambda$10330.14 line is visible (Figures \ref{fig:hband} and \ref{fig:entire}).
While Sr is more ionized due to the high atmospheric temperature in this star ($\teff = 12000$\,K), its high abundance of ${\rm [\ion{Sr}{ii}/H]}=1.40$ \citep[Table~\ref{tab:ews}; cf. ][]{Sadakane1992} allows the strongest lines to remain visible.

The strongest $H$-band \ion{Ce}{iii} lines, $\lambda\lambda$15851.88, 15961.16, and 16133.17, are detected in all targets except HD~8441, HD~182564, and HD~190165 (Figures \ref{fig:hband} and \ref{fig:entire}).
In these three stars, the lines are not clearly visible either because the Ce abundance is lower than literature value or because the absorption lines are broadened due to relatively high rotational velocities.
In addition, more than 10 \ion{Ce}{iii} lines are detected across the observed wavelength range in most ApBp stars: 
$\lambda\lambda$11071.35, 11094.40, 12353.17, 12760.44, 12825.13, 12926.64, 13155.11, 13342.83, 15098.51, 15720.13, 15964.93, and 16292.64. 
The excitation energies of these detected \ion{Ce}{iii} lines are $\lesssim 1$\,eV. The $\lambda$15720.13 line, a candidate for the 1.45\,$\umu$m features in the AT2017gfo spectra \citep[the 1.58\,$\umu$m features in their definition; see Figure~\ref{fig:kn}]{Gillanders2024}, is detected in four stars, suggesting that it may contribute to kilonova spectra but secondary to the three strongest lines.

The \ion{Ce}{iii} $\lambda$12760.44 and $\lambda$12353.17 lines are candidates for the 1.15\,$\umu$m features in the AT2017gfo spectra \citep[the 1.23\,$\umu$m features in their definition]{Gillanders2024}. 
The former is detected in HD~5797, HD~22316, and HD~176232, while the latter is detected in HD~965, HR~465, and HD~188041. 
However, the region around 12760\,{\AA} in most stars was masked due to the overlap with sky O$_2$ lines depending of radial velocities and line widths. 
Checking the unmasked data revealed that the $\lambda$12760.44 line is present in four additional stars, including three exhibiting the $\lambda$12353.17 line as well \citep[see also][]{Tanaka2023}. 
This indicates that both lines are rather strong, with the $\lambda$12760.44 being the primary. 
Conversely, the ground-state $\lambda$10723.27 line, one of candidates for the 0.9\,$\umu$m features in the AT2017gfo spectra \citep[the 1.08\,$\umu$m features in their definition]{Gillanders2024}, is not detected in any of the sample stars. This indicates that the transition probability of this line is small, and it is unlikely that this line contributes significantly to kilonova spectra.

\citet{Rahmouni2025} highlighted the importance of \ion{Gd}{iii} lines for kilonova spectra, identifying $\lambda$14336.79 and $\lambda$17479.48 lines in HR~465. We confirm the presence of both \ion{Gd}{iii} lines in our HR~465 spectrum and detect them in five additional stars: HD~965, HD~22316, HD~176232, HD~188041, and HD~203932 (Figure~\ref{fig:entire}). 
Although the $\lambda$14336.79 line is located in a masked region due to severe telluric absorption, knowing its wavelength allowed us to force a measurement of its EW.
The EWs of both \ion{Gd}{iii} lines exceed those of the \ion{Sr}{ii} and \ion{Ce}{iii} lines in some stars (Figure~\ref{fig:Ap}). This confirms that the \ion{Gd}{iii} lines are also intrinsically strong and likely relevant to kilonova spectra.

Among doubly ionized lanthanides, \ion{Nd}{iii} is the most frequently detected species after \ion{Ce}{iii} and \ion{Gd}{iii}, appearing in HD~965, HD~203932, and HR~465. 
The \ion{Nd}{iii} absorption lines in the HR~465 spectrum are much stronger than that in \citet{Tanaka2023}, likely due to the abundance variability (see Figure~\ref{fig:Ap} and their figure 4 for comparison). The number of \ion{Nd}{iii} lines is intrinsically larger than that of \ion{Ce}{iii} because of a more complex electron structure, as seen in HR~465. The \ion{Nd}{iii} lines are distributed across the observed wavelength range, with excitation energies of $\approx1.4$--2\,eV.

Absorption lines of other lanthanides, including \ion{Ce}{ii}, \ion{Pr}{ii}, \ion{Pr}{iii}, \ion{Nd}{ii}, \ion{Sm}{ii}, \ion{Eu}{ii}, \ion{Gd}{ii}, \ion{Tb}{iii}, \ion{Dy}{ii}, \ion{Ho}{ii}, \ion{Ho}{iii}, \ion{Er}{ii}, and \ion{Tm}{ii} ($Z=58$--60, 62--69), are also detected mainly in the $YJ$ bands of HD~965 and HR~465 (Figure~\ref{fig:entire}). 
More lines of singly ionized lanthanides are detected as compared to doubly ionized ones, which is natural because of more complex atomic structures.
Among singly ionized lanthanides, most lines have comparable strength and none stands out significantly. Notably, the lines of \ion{Tb}{ii} and \ion{Tm}{ii} listed as potential candidates for the AT2017gfo spectral features by \citet{Gillanders2024} are not detected in our sample.

An important exception among singly ionized species is \ion{Eu}{ii}. 
Its lines, including $\lambda\lambda$9885.83, 9900.98, 9991.33, 10022.27, 10036.94, 10068.78, 10145.69, 10168.35, 10311.80, and 15075.20, are detected in five ApBp stars (HD~965, HD~5797, HD~188041, HD~203932, and HR~465; Figures \ref{fig:yband} and \ref{fig:entire}), where they are sometimes comparable to or stronger than the \ion{Sr}{ii} triplet. 
The excitation energies of these detected lines are $\approx2.1$\,eV. 
The reason why \ion{Eu}{ii} lines are easily detected, as in optical spectrum, is that its atomic structure is relatively simple, involving a half-filled 4f-shell and an additional electron in another orbital, similar to \ion{Gd}{iii} \citep{Rahmouni2025}. 
While many \ion{Eu}{ii} lines are listed as potential candidates for the AT2017gfo spectral features by \citet{Gillanders2024}, the $\lambda$10311.80 and $\lambda$15075.20 lines are detected only in one and two stars, respectively, and $\lambda\lambda$10738.70, 10906.58, and 15321.90 lines are not detected in our sample. This suggests that these lines are rather weak and unlikely to be important for kilonova spectra.

In addition to species discussed above, we mention constraints on \ion{La}{iii} and \ion{Y}{ii}. 
Prominent \ion{La}{iii} transitions lie at 13898.27, 14100.04 and 17882.98\,{\AA} \citep{Domoto2022}, with the former two previously identified as key candidates for the $1.3\,\umu$m feature in kilonova spectra. In our spectra, the wavelength range around 14000\,{\AA} is obscured by severe telluric absorption, while 17883\,{\AA} falls outside the IRD spectral range. Other \ion{La}{iii} lines within our coverage originate from high-excitation energy levels \citep{Floers2026} and are thus too weak for detection. Our non-detection therefore does not imply that \ion{La}{iii} is unimportant for kilonova spectra, but rather reflects the constraints imposed by telluric and instrumental coverage limits.
In contrast, several \ion{Y}{ii} lines are detected in our spectra, \eg at 10108.29\,{\AA} and 10248.03\,{\AA} (see Figure~\ref{fig:entire}). While these NIR detections do not directly measure the optical features discussed in kilonova literature \citep[around 700\,nm;][]{Sneppen2023}, it is consistent with the idea that \ion{Y}{ii} optical transitions can be important for kilonova spectra.

Figures \ref{fig:Ap} and \ref{fig:Am} show the EWs of the top 50 strongest absorption lines in the sample ApBp and AmFm stars, respectively. 
When ranking by EW, likely blended lines are excluded and only single lines are presented. 
The excluded lines account for only a few percent of the detected lines, although sometimes strong lines are excluded; for example, the \ion{Sr}{ii} $\lambda$10917.88 line, one of the triplet, blends with strong \ion{Mg}{ii} lines in stars with relatively high rotational velocities. 
Including blended lines marginally affects the ranking, mainly in AmFm stars, but does not affect our conclusions. HD~190165 is not included in Figure~\ref{fig:Am} because its high rotational velocity causes severe line blending. 
Notably, all of the strongest lines that remained unidentified in Figure~4 of \citet{Tanaka2023} are identified here, including \ion{Mg}{i}, \ion{Si}{i}, \ion{Si}{ii}, and \ion{Fe}{ii}, as well as \ion{Gd}{iii} \citep{Rahmouni2025} and \ion{Nd}{iii}. 
Neutral lines were missed in that work because they were not included in the line-identification procedure, while some singly ionized lines were missed because their high excitation energies exceeded the matching criteria used for line identification. The identification of \ion{Nd}{iii} lines is enabled by the list of \citet{Floers2026}.

Overall, Sr and lanthanides dominate among heavy elements in all stars. While some lines remain unidentified, none are significantly stronger than, \eg \ion{Sr}{ii} or \ion{Ce}{iii} lines. Because our line identification relies on completeness of the experimental data, especially for low-lying energy levels, unidentified lines may originate from unknown low-lying levels or higher excitation states of heavy elements. 
Even if the low-lying data are incomplete, the observed lines remain not very strong in our CP-star spectra. Furthermore, due to the lower temperatures in kilonovae, any high-excitation transition will be even more suppressed in kilonova spectra.
Consequently, contribution of these unidentified lines to kilonova spectra is negligibly minor.

\section{Discussions}
\label{sec:discussion}
Most of our CP-star spectra exhibit the \ion{Sr}{ii} triplet and the $H$-band \ion{Ce}{iii} lines. In addition to more than 10 \ion{Ce}{iii} lines, strong absorption lines of \ion{Gd}{iii}, \ion{Eu}{ii}, and \ion{Nd}{iii} are detected, with the number of stars in which these lines are detected decreasing in the order of \ion{Gd}{iii}, \ion{Eu}{ii}, and \ion{Nd}{iii}. 
Figure~\ref{fig:kn} compares the spectra of AT2017gfo \citep{Pian2017, Smartt2017, Tanvir2017} with the strong transitions of these ions. 
Lines that are ranked in the top 50 for EWs in at least two or three individual stars are shown with thin and thick bars, respectively (Figure~\ref{fig:Ap}--\ref{fig:Am}), with their rest wavelengths blueshifted by $v=0.1\,c$.
The 0.9\,$\umu$m features in the AT2017gfo spectra are consistent with the \ion{Sr}{ii} triplet (at $v=0.2c$) \citep{Watson2019}, but also closely match the \ion{Eu}{ii} lines. 
Similarly, the 1.45\,$\umu$m features in the AT2017gfo spectra well agree with the \ion{Ce}{iii} lines \citep{Domoto2022} and some \ion{Nd}{iii} lines.

To discuss whether \ion{Eu}{ii} lines may compete with the \ion{Sr}{ii} triplet, we show the measured EWs of the \ion{Sr}{ii} and \ion{Eu}{ii} lines as a function of their abundance ratio in the left panel of Figure~\ref{fig:ew}. 
The transition wavelength, excitation energy, and $\log gf$ based on the VALD line list for each transition are given in the legend. 
Hereafter, \eg \ion{Eu}{ii} abundance denotes the Eu abundance measured using \ion{Eu}{ii} lines. 
We adopt mean \ion{Eu}{ii} abundances from the literature regardless of the model-data agreement, whereas the \ion{Sr}{ii} abundances are estimated in this work (Section~\ref{sec:ana}). The plotted data are provided in Table~\ref{tab:ews}.
The measured EWs increase approximately linearly with abundance. 
The observed scatter is expected because the strength of absorption lines depends not only on abundances but also on stellar atmospheric parameters. 
The theoretical behaviour of EWs with varying atmospheric parameters (\ie the curve of growth) is briefly discussed in Appendix \ref{sec:cog}.
Systematic uncertainties in $\teff$, $\log g$, and abundance ratios in dwarf stars, generally caused by spectral analysis, can be as large as $\pm200$\,K, $\pm0.3$\,dex, and $\pm0.5$\,dex, respectively \citep{Hinkel2016}. The persistence of the linear trend despite such variations and uncertainties indicates that the relation is robust.

The strength of the \ion{Eu}{ii} lines become roughly comparable to that of the \ion{Sr}{ii} triplet when ${\rm [Eu/H]}$ exceeds ${\rm [Sr/H]}$ by roughly 2--3\,dex, \ie when their absolute abundances are comparable. This is because their ionization fractions are expected to be comparable under LTE due to similar ionization potentials of $\approx5.7$\,eV and because the excitation energies and $gf$-values are also similar among these lines. 
Observations of metal-poor $r$-process enhanced stars, which are thought to reflect intrinsic nucleosynthesis outcome on their surfaces, suggest that the typical Eu abundance is at least an order of magnitude lower than Sr abundance \citep[e.g.,][]{Roederer2022}. Nevertheless, in neutron star mergers, Eu can be produced by a factor of a few or more than Sr in neutron-rich ejecta \citep[grey dashed line in Figure~\ref{fig:abun}; see, e.g.,][]{Fernandez2017, Fujibayashi2023, Wanajo2024}. 
If a highly heavy-element-enriched abundance pattern is realized in strongly neutron-rich ejecta, Eu may contribute to the kilonova spectral features along with Sr. In such a case, because the rest wavelengths of the strong \ion{Eu}{ii} lines are shorter than those of the \ion{Sr}{ii} triplet, we speculate that resulting spectral features would appear more blueshifted than what was observed in AT2017gfo.

Similarly, to discuss whether \ion{Nd}{iii} lines may compete with the \ion{Ce}{iii} lines, we show the EWs of the \ion{Ce}{iii} and \ion{Nd}{iii} lines as a function of abundance ratio in the right panel of Figure~\ref{fig:ew}, with the data are provided in Table~\ref{tab:ews}. Note that the $gf$-values of the \ion{Ce}{iii} and \ion{Nd}{iii} lines shown in the legend are taken from \citet{Domoto2023} and \citet{Floers2026}, respectively, and are experimentally uncertain. 
HD~965 and HD~203932, which are known to exhibit the REE anomalies, are excluded from the plot because only the \ion{Ce}{ii} abundance is available in the literature. 
The linear trend persists despite uncertainties in stellar parameters, again indicating a robust relation (see also Appendix \ref{sec:cog}).

The strength of the \ion{Nd}{iii} lines become comparable to that of the \ion{Ce}{iii} lines only when Nd is roughly 100 times more abundant than Ce. 
This can be understood from the higher excitation energies (\ie smaller level population for a given temperature) of the \ion{Nd}{iii} lines. Also, although the $gf$-values are uncertain, available data indicate that those of the \ion{Nd}{iii} lines are smaller than the \ion{Ce}{iii} lines. 
Thus, an extreme abundance pattern within lanthanides would be required for \ion{Nd}{iii} lines to dominate over \ion{Ce}{iii} lines in kilonova spectra. 
However, such an abundance pattern is unlikely, as two elements separated by an atomic number of only two.
Nd is therefore less likely to contribute comparably to Ce in kilonova spectra.

On the other hand, \ion{Nd}{iii} also exhibits several strong transitions close to the \ion{Sr}{ii} triplet. The behaviour of their EWs is similar to that in the right panel of Figure~\ref{fig:ew}. 
As with \ion{Eu}{ii}, if a highly heavy-element-enriched abundance pattern is realized in neutron-rich ejecta of neutron star mergers, \ion{Nd}{iii} may still contribute to kilonova spectra along with the \ion{Sr}{ii} and \ion{Eu}{ii} lines, or might even dominate over them at high temperatures due to its slightly lower ionization potential. 
In such a case, because many \ion{Nd}{iii} lines with comparable strength are distributed across the investigated wavelengths, multiple spectral features would be expected. 
Alternatively, Nd ion lines could simply dominate the global opacity, shaping the overall spectrum due to large number of transitions \citep{Even2020, Tanaka2020, Fontes2026}.

\section{Summary}
\label{sec:conclusion}
We have presented high-resolution, NIR $Y$-, $J$-, and $H$-band spectra of 14 CP stars covering a range of $\teff$, $\log g$, and abundance ratios, obtained with Subaru/IRD. 
This sample represents the largest high-resolution NIR multi-band spectral dataset of CP stars assembled to date. As a result, numerous absorption lines of heavy elements have been detected at NIR wavelengths for the first time using the latest line list \citep{Floers2026}, expanding our empirical knowledge of strong heavy-element transitions.
The stellar parameters covered by our sample bridges regimes where doubly ionized species dominate to where singly and doubly ionized species coexist, effectively mirroring the physical conditions expected in kilonova ejecta a few days after mergers. This allows comprehensive study of heavy-element transitions for a better understanding of the absorption features in kilonova spectra.

Most of the sample spectra exhibit the \ion{Sr}{ii} triplet in the $Y$ band and the \ion{Ce}{iii} lines in the $H$ band, confirming the intrinsic importance of these transitions as highlighted in previous work \citep{Watson2019, Domoto2022, Domoto2023, Tanaka2023}. 
In addition to the strong \ion{Gd}{iii} lines \citep{Rahmouni2025}, we identified \ion{Eu}{ii} and \ion{Nd}{iii} lines in several stars. 
For the specific \ion{Ce}{iii} and \ion{Eu}{ii} lines proposed by \citet{Gillanders2024} as candidates for the spectral features in AT2017gfo, we confirmed certain detections while establishing non-detection for others, providing the first observational verification.
Furthermore, we showed that if a highly heavy-element-enriched abundance pattern is realized in strongly neutron-rich ejecta of neutron star mergers, \ion{Eu}{ii} and \ion{Nd}{iii} may contribute to the spectral features comparably to \ion{Sr}{ii}.
Importantly, we found no unidentified lines systematically stronger than these transitions even with the increased number of stars in our sample. These results strengthen the conclusion that the \ion{Sr}{ii} and \ion{Ce}{iii} transitions play the primary role in producing absorption features in kilonova spectra, and provide new insights into the potential contributions of \ion{Eu}{ii} and \ion{Nd}{iii}.

Finally, this study establishes a fundamental data important for analysing future kilonova observations. While our current understanding of kilonovae relies on a single event AT2017gfo, future kilonovae are expected to exhibit different spectra, reflecting diversity in ejecta properties and abundances. Our results thus provide a necessary baseline to decode such diversity. In addition, for the construction of atomic data, the results provide immediate targets for laboratory measurements. This will guide and encourage future atomic physics studies, helping to systematically improve the accurate line list required for analysing astrophysical spectra.

\section*{Acknowledgements}
We thank Y. Kasagi for the support to use \textsc{PyIRD}. This work is based on data collected at the Subaru Telescope, which is operated by the National Astronomical Observatory of Japan. We are honoured and grateful for the opportunity of observing the Universe from Maunakea, which has the cultural, historical, and natural significance in Hawaii. This work was supported by the Grant-in-Aid for JSPS Fellows (grant No. 25KJ0075), the Grants-in-Aid for Scientific Research from JSPS (grant Nos. 21H04997, 23H00127, 23H04891, 23H04894, 23H05432, 25K01046, 26H01433, 26K21726), the JST FOREST Program (grant No. JPMJFR212Y), and NIFS Collaborative Research Program (grant Nos. NIFS22KIIF005, NIFS24KIIQ013). S.R. acknowledges support from the Graduate Program on Physics for the Universe (GP-PU) at Tohoku University.

\section*{Data Availability}
The raw spectral data used in this article are available for download from the SMOKA archives.\footnote{\url{https://smoka.nao.ac.jp/}}



\bibliographystyle{mnras}
\bibliography{references} 



\appendix

\section{New transition wavelengths of \ion{Sr}{ii}}
\label{sec:sr}
During the analysis, we detect unidentified absorption lines at wavelengths close to known \ion{Sr}{ii} transitions. These lines are detected at consistent wavelengths across different spectra, and some appear in more than half of the spectra, suggesting that the element responsible for the absorption is common. The information of the \ion{Sr}{ii} lines near the detected unidentified features is summarised in Table~\ref{tab:sr}. The vacuum transition wavelengths are those expected from the energy levels listed in the NIST ASD \citep{nist}, and the $gf$-values are adopted from \citet{Kurucz2017} as in the latest VALD line list.

The differences between the wavelengths expected for \ion{Sr}{ii} and those detected in our spectra are as large as 0.8\,{\AA}. This discrepancy is larger than the typical accuracy of wavelength calibration in our data. On the other hand, the NIST ASD indicates that the experimental accuracy of the \ion{Sr}{ii} data is limited: the transition wavelengths are uncertain by up to 0.5\,{\AA}, and the energy levels by up to 0.3\,cm$^{-1}$. In fact, in model spectra with the Sr abundances that reasonably reproduce the triplet, the predicted strength of the \ion{Sr}{ii} lines are comparable to that of the nearby unidentified features in the observed spectra. Thus, the detected lines are most likely due to \ion{Sr}{ii}. Our results provide improved wavelengths and highlight the limitations in the accuracy of \ion{Sr}{ii} levels at relatively high excitation. Nevertheless, since the lower energies of these transitions are $> 5$\,eV, they are not likely to play a significant role in kilonova spectra.

\begin{table*}
\begin{threeparttable}
	\centering
	\caption{\ion{Sr}{ii} transitions with new transition wavelengths.}
	\label{tab:sr}
	\begin{tabular}{ccccccccc}\hline
	$\lambda_{\rm vac}$$^{\rm a}$ ({\AA})  & $\log gf$$^{\rm b}$   & Lower level &  $E_{\rm lower}$ (cm$^{-1}$) &  Upper level &  $E_{\rm upper}$ (cm$^{-1}$) & $\lambda_{\rm vac, obs}$$^{\rm c}$ ({\AA}) & $\Delta\lambda$$^{\rm d}$ ({\AA}) & Number of detection \\ \hline
	12017.3  & 0.322      &  6s $^2{\rm S}_{1/2}$ &  47736.53  & 6p $^2{\rm P}_{3/2}$   & 56057.9    &12017.9  &   0.6     & 9 \\
	12448.4  & 0.017      &  6s $^2{\rm S}_{1/2}$ &  47736.53  & 6p $^2{\rm P}_{1/2}$  & 55769.7    &12449.2  &   0.8     & 9  \\
	12978.5  & 0.567      &  5d $^2{\rm D}_{3/2}$ &  53286.31  & 4d $^2{\rm F}_{5/2}$  & 60991.34   &12978.0  &  $-$0.5 & 4 \\
	13126.2  & $-$0.584 &  5d $^2{\rm D}_{5/2}$ &  53372.97 & 4d $^2{\rm F}_{5/2}$  & 60991.34   &13125.9  &  $-$0.3 & 3 \\
	13128.4  & 0.717      &  5d $^2{\rm D}_{5/2}$ &   53372.97 & 4d $^2{\rm F}_{7/2}$  & 60990.04   &13127.9  &  $-$0.5 & 4 \\
	15213.5  & 0.009       &  4f $^2{\rm F}_{7/2}$ &  60990.04 &  6d $^2{\rm D}_{5/2}$ & 67563.15   &15214.3  &   0.8     &  3 \\
	15216.5  & $-$1.293 &  4f $^2{\rm F}_{5/2}$ &  60991.34  &  6d $^2{\rm D}_{5/2}$  & 67563.15   &     -         &       -     & 0 \\ 
	15310.4  & $-$0.149 &  4f $^2{\rm F}_{5/2}$ &  60991.34  &  6d $^2{\rm D}_{3/2}$  & 67522.87   &15310.9  &   0.5     &  1 \\\hline
	\end{tabular}
	\begin{tablenotes}
	\item[a] Vacuum transition wavelength of \ion{Sr}{ii} expected from the energy level differences listed to the next columns \citep{nist}. The energy levels have the uncertainties of 0.01--0.3\,cm$^{-1}$.
	\item[b] $gf$-value listed in the VALD line list \citep{Kurucz2017}.
	\item[c] Central wavelength of the detected absorption line.
	\item[d] Difference between the expected and detected wavelengths.
	\end{tablenotes}
\end{threeparttable}
\end{table*}

\section{Comparison between observed and modelled equivalent widths}
\label{sec:cog}
Table~\ref{tab:ews} summarises the observed EWs of the selected lines and the mean abundances of the corresponding ions. These data are plotted in Figure~\ref{fig:ew} except for \ion{Gd}{iii}. 
For stars known to exhibit the REE anomalies where literature abundances are available only for a different ionization state, the values are listed in italics; the \ion{Ce}{iii} data for such stars are not plotted in Figure~\ref{fig:ew}.
For HD~225914, HD~187254, HD~190165, and HD~225463, only total elemental abundances are available, which we adopt here. For the remaining Am stars, only abundances of singly-ionized ions are available, but we simply adopt them given the absence of the REE anomalies.

\begin{figure*}
	\includegraphics[width=\linewidth]{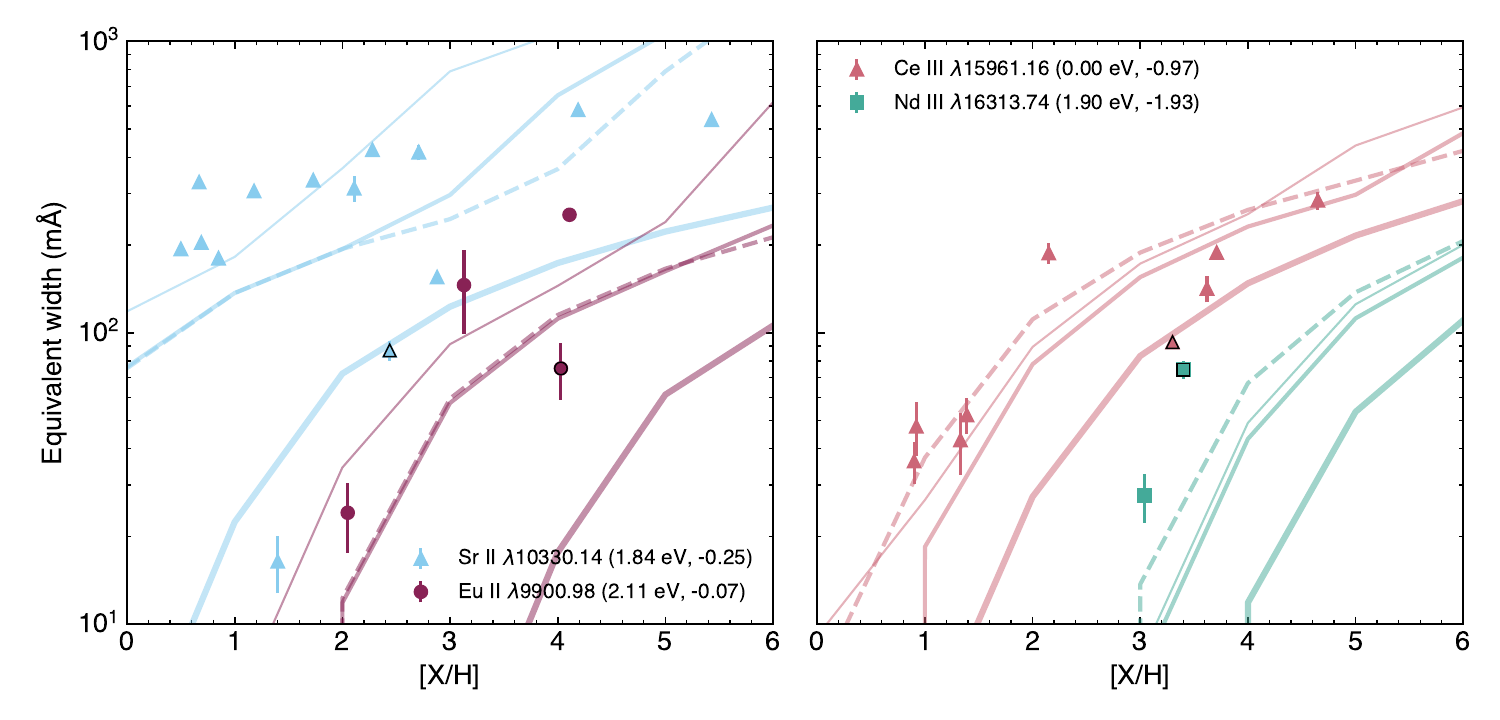}
	\caption{Theoretical EWs of the selected lines as a function of abundance, overplotted with the observed values. Solid curves show models of different temperatures, $\teff=7400$, 8670, and 11000\,K (thin to thick) with a fixed $\log g=4.3$. Dashed curves show models with $\teff=8670$\,K and $\log g=3.4$. All models are calculated assuming ${\rm [Fe/H]}=0.5$, $\xi=1.5$\,\kms, and $v \sin i = 2.0$\,\kms, using the solar abundance, except for the target elements enhanced by a factor of 0 to +6 dex.
	}
	\label{fig:cog}
\end{figure*}

To better understand the observed scatter in Figure~\ref{fig:ew}, Figure~\ref{fig:cog} shows the behaviour of the theoretical EWs for the selected lines as a function of abundance (\ie the curve of growth). The solid curves represent models with $\teff=7400$, 8670, and 11000\,K at a fixed $\log g=4.3$ (thin to thick) to see the temperature dependence. The dashed curves represent models with $\teff=8670$\,K and $\log g=3.4$ to see the pressure dependence. These synthetic tracks are calculated using \textsc{Turbospectrum} through \textsc{iSpec} as in Section~\ref{sec:ana}, assuming fixed values of ${\rm [Fe/H]}=0.5$, $\xi=1.5$\,\kms, and $v \sin i=2.0$\,\kms. Abundance patterns are fixed to be the solar values, except for the target elements enhanced by 0 to $+$6\,dex. The modelled EWs are measured as done for the observed spectra (Section~\ref{sec:ana}).

The modelled curve of growth broadly agree with the observed trends. For all lines, the EW decreases at higher temperatures for a given ${\rm [X/H]}$, particularly pronounced for \ion{Sr}{ii} and \ion{Eu}{ii} lines. This is primarily because the mean ionization state shifts toward higher states with increasing temperature \citep{Gray2005}. For \ion{Sr}{ii} $\lambda$10330.14 line at $\teff<9000$\,K, the line becomes saturated at ${\rm [Sr/H]}>2$. The EW increases at larger $\log g$ for a given ${\rm [Sr/H]}$ due to pressure broadening in the line wings. Conversely, for \ion{Ce}{iii} $\lambda15961.16$ and \ion{Nd}{iii} $\lambda$16313.74 lines, a lower $\log g$ enhances the EWs for a given ${\rm [X/H]}$ due to the pressure effect on ionization states. Overall, these models confirm that the observed scatter can be attributed to variations in the atmospheric parameters of the samples stars.

Note, however, that several uncertainties affect EWs. The deviations of the observed EWs from the modelled tracks, particularly seen in the \ion{Eu}{ii} and \ion{Nd}{iii} lines, are likely caused by uncertainties in the $gf$-values, the literature abundances, or both. Because detailed measurements of their $gf$-values and abundances are beyond the scope of this work, our discussion relies directly on empirical observational results rather than models, to avoid being biased by these systematic uncertainties.

\begin{landscape}
\begin{table}
	\caption{Observed EWs (in units of m{\AA}) of the selected lines. They are plotted in Figure~\ref{fig:ew} except for \ion{Gd}{iii}.
		     Mean ionic abundances used in the figure are also listed, based on this work for Sr (Section~\ref{sec:ana}) and the literature for others.
		     Italic font indicates abundance derived using different ionization states in stars where the REE anomalies are observed.
		     Blanks for abundances indicate where no measurements are available, and dashes (-) for EWs represent non-detection.}
	\label{tab:ews}
	\begin{tabular}{ccccccccccccccc}\hline
		& HD~965  & HD~5797 & HD~8441 & HD~22316 & HR~465 & HD~176232 & HD~188041 & HD~203932 & HD~225914 & HD~182564 & HD~187254 & HD~189849 & HD~190165 & HD~225463 \\ \hline
		[\ion{Sr}{ii}/H] & 2.71 & 5.43 & 2.88 & 1.40 & 2.44 & 0.50 & 4.19 & 0.85 & 2.11 & 0.69 & 1.18 & 0.67 & 2.28 & 1.73 \\
		$\lambda$10039.41 & 194 & 240 &    62 &  -  & 35 &  74  & 239 &  52  &  74  &  38  & 113 & 141 & 244 & 152 \\
		$\lambda$10330.14 & 416 & 538 & 155 & 16 & 87 & 194 & 582 & 180 & 313 & 204 & 306 & 329 & 319&  334 \\ \hline
		[\ion{Eu}{ii}/H] &        & 3.13 & 2.16 & 3.28 & 4.03 & 1.88 & 4.11 & 2.05 & 1.93 & 1.44 & 0.93 & 1.84 & & 1.34 \\
		$\lambda$9900.98   & 184 & 146 & - & - & 75 & - & 254 & 24 & - &  - &  - & - & - &  - \\
		$\lambda$9991.38   & 166 &  70  & - & - & 93 & - & 264 &  -   & - &  - &  - & - & - & - \\
		$\lambda$10022.27 & 199 & 137 & - & - & 61 & - & 273 & 33 & - & -  &  - & - & - & - \\ \hline
		[\ion{Ce}{iii}/H] & {\it 2.48} & 2.15 & 0.76 & 3.62 & 3.30 & 3.71 & 4.65 & {\it 1.19} & 0.92 & 1.81 & 1.33 & 0.90 & 1.33 & 1.39 \\
		$\lambda$15851.88 & 354 & 220 & - & 161 & 101 & 196 & 326 & 241 &  67  & - & 46 & 34 & - & 51 \\
		$\lambda$15961.16 & 356 & 188 & - & 142 &  93  & 188 & 284 & 161 &  48  & - & 43 & 36 & - & 52 \\
		$\lambda$16133.17 & 433 & 139 & - & 151 &  78  & 190 & 306 & 176 & 109 & - & 60 & 23 & - & 58 \\ \hline
		[\ion{Nd}{iii}/H] & & 2.13 & 1.36 & & 3.40 & 3.40 & 2.75 & 3.04 & 0.90 & 0.69 & 1.18 & 0.67 & 2.28 & 1.73 \\
		$\lambda$15231.35 &  66  & - & - & - & 83 & - & - & 19 & - & - & - & - & - & - \\
		$\lambda$16183.57 & 141 & - & - & - & 85 & - & - & 52 & - & - & - & - & - & -\\
		$\lambda$16313.74 & 161 & - & - & - & 68 & - & - & 28 & - & - & - & - & - & - \\ \hline
		[\ion{Gd}{iii}/H] & & {\it 2.57} & 1.70 & {\it 4.48} & 3.20 & 3.11 & {\it 3.49} & {\it 1.81} & 1.93 & 1.26 &  & 0.95 &  &  \\
		$\lambda$14336.79 & 264 & - & - & 286 & 122 &  87  & 298 & 19 & - & - & - & - & - & - \\
		$\lambda$17479.48 & 386 & - & - & 237 & 125 & 152 & 648 & 52 & - & - & - & - & - & -\\ \hline
	\end{tabular}
\end{table}
\end{landscape}

\section{Entire IRD spectra of CP stars}
\label{sec:entire}
Figure~\ref{fig:entire} shows the entire NIR spectra of our sample CP stars (object names are shown in each panel), obtained with Subaru/IRD in the $Y$, $J$, and $H$ bands. 
Grey shaded regions without line identification indicate wavelength ranges affected by strong telluric absorption. Absorption lines with EWs $> 9$\,m{\AA} for HD~8441 and $> 16$\,m{\AA} for others are marked by vertical bars with species names (colours: heavy elements, black: unidentified, and grey: others), and false of absorption lines, \ie residual of telluric correction, are also marked by light-grey bars.

\begin{figure*}
	\includegraphics[width=\linewidth]{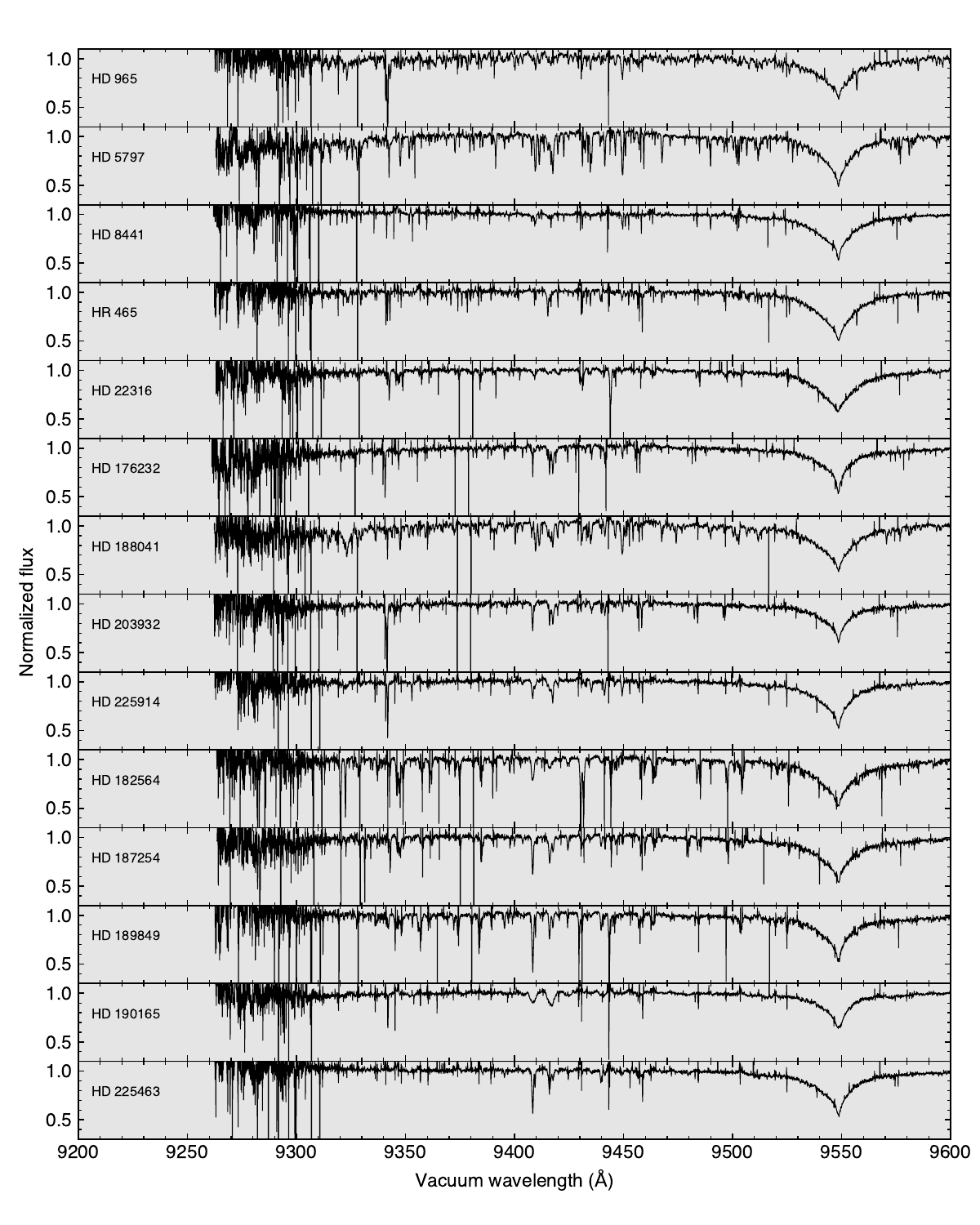}
	\caption{Normalized spectra of 14 CP stars in the $YJH$ bands obtained with Subaru/IRD.
}
	\label{fig:entire}
\end{figure*}

\foreach \n in {2,3,4,5,6,7,8,9,10,11,12,13,14,15,16,17,18,19,20,21,22}{
	\begin{figure*}\ContinuedFloat
	\centering
	\includegraphics[width=\textwidth,page=\n]{fig_c1.pdf}
	\contcaption{Normalized spectra of 14 CP stars in the $YJH$ bands obtained with Subaru/IRD.}
\end{figure*}
}


\bsp	
\label{lastpage}
\end{document}